\documentclass[12pt]{iopart}

\pdfoutput=1

\usepackage{amssymb}
\usepackage{graphicx}
\usepackage{verbatim}
\usepackage{color}
\usepackage[numbers,sort&compress]{natbib}

\graphicspath{{figures/}{/}}

\newcommand {\avg}[1]      {\langle#1\rangle}

\newcommand {\nup}         {N_{\mathord{\uparrow}}}

\newcommand {\TrB} {\Tr_\mathrm{B}\nolimits}
\newcommand {\binom}[2] {{{#1} \choose {#2}}}

\newcommand  {\ket}[1]      {\left|#1\right\rangle}
\newcommand  {\bra}[1]      {\left\langle#1\right|}
\newcommand  {\braket}[2]   {\left\langle#1\mid#2\right\rangle}

\newcommand  {\abs}[1]      {|#1|}

\newcommand{\nb}{N_\mathrm{b}}

\begin{document}

\title[Global characteristics of all eigenstates of local many-body Hamiltonians: \ldots]{Global characteristics of all eigenstates of local many-body Hamiltonians: participation ratio and entanglement entropy}

\date{\today}

\author{W Beugeling, A Andreanov and Masudul Haque}

\address{Max-Planck-Institut f\"ur Physik komplexer Systeme, N\"othnitzer Stra\ss e 38, 01187 Dresden, Germany} 


\begin{abstract}
	In the spectrum of many-body quantum systems, the low-energy eigenstates were the traditional focus of research. The interest in the statistical properties of the full eigenspectrum has grown more recently, in particular in the context of non-equilibrium questions.  
Wave functions of interacting lattice quantum systems can be characterized either by local observables, or by global properties such as the participation ratio (PR) in a many-body basis or the entanglement between various partitions.
We present a study of the PR and of the entanglement entropy (EE) between two roughly equal spatial
partitions of the system, in all the eigenfunctions of local Hamiltonians. Motivated by the
similarity of the PR and EE --- both are generically larger in the bulk and smaller near the edges
of the spectrum --- we quantitatively analyze the correlation between them.
We elucidate the effect of (proximity to) integrability, showing how low-entanglement and low-PR states appear also in the middle of the spectrum as one approaches integrable points.  We also determine the precise scaling behavior of the eigenstate-to-eigenstate fluctuations of the PR and EE with respect to system size, and characterize the statistical distribution of these quantities near the middle of the spectrum.
\end{abstract}


\maketitle

\section{Introduction}

Non-equilibrium dynamics of thermally isolated quantum systems has enjoyed substantial recent interest. An isolated system has no relaxation mechanism toward the low-lying parts of the many-body spectrum. Thus, eigenstates far from the edges of the spectrum may be more important for a non-equilibrium experiment than the low-energy parts of the spectrum, which is the traditional focus of interest of many-body quantum theory. Given this context, the properties of the \emph{full} eigenspectrum of many-body interacting quantum systems has become important and of interest to a growing community.

Many-body eigenstates can be characterized in various ways.  In connection to questions involving thermalization, the statistical behavior of various \emph{local} observables in eigenstates has been widely studied, both in non-integrable~\cite{RigolSrednicki2012, KhatamiEA2012,
  BiroliEA2010, Rigol2009PRL,Rigol2009PRA, GramschRigol2012, HeEA2013} and integrable
systems~\cite{IkedaEA2013, Alba_arXiv1409}, and also in comparisons between the two
cases~\cite{RigolEA2008, NeuenhahnMarquardt2012, SteinigewegEA2013, BeugelingEA2014PRE, SorgEA_PRA2014}.
In particular, the eigenstate thermalization hypothesis (ETH) states that the mechanism by which non-integrable (`chaotic' or `generic') systems thermalize is the smooth behavior of the eigenstate expectation values of local observables as a function of eigenenergies~\cite{Deutsch1991,Srednicki1994,RigolEA2008}.

Clearly, local observables are not the only properties of many-body eigenstates that are of interest. In this work, we focus on two global characteristics of such eigenstates. The \emph{entanglement} between two spatial partitions of a system is now well-appreciated as a useful characterization of quantum correlations in many-body wave functions.
In addition, wave functions can be characterized by their (inverse) participation ratio in some basis, which quantifies the number of basis states that take part in forming this wave function, i.e., the amount of Hilbert-space delocalization in this basis. A natural choice is to use the basis of real-space configurations.  The participation ratio (PR) in this basis provides a generalization, for many-body systems, of the concept of delocalization in single-particle physics.  In recent years the PR (or its inverse, the IPR) in various many-body bases has been widely used to characterize delocalization in many-body Hilbert spaces~\cite{BerkovitsAvishai_PRL98, BrownEA2008,   RigolSantos2010,SantosRigol2010, YurovskyOlshanii2011, KotaEA2011, BuccheriEA2011, CanoviEA2011, SantosEA2012PRE, BrandinoEA2012, NeuenhahnMarquardt2012, CanoviEA2012, ZangaraEA2013, HeEA2013, TorresHerreraSantos2013, ZiraldoSantoro2013, SerbynPapicAbanin_PRL2013b, BauerNayak_JSTAT2014}, e.g., in the context of many-body localization or in the context of ETH studies.
Intuitively, one would expect that an eigenstate with large PR in this basis would also have large entanglement between two spatial parts of the system.

In this work, we explore both the entanglement between nearly equal-sized real-space partitions, as measured using the (von Neumann) entanglement entropy (EE), and the PR in the basis of real-space configurations, of \emph{every} eigenstate of interacting many-body systems.  We show how these quantities behave in the bulk of the spectrum as opposed to the edges of the spectrum. Tuning the many-body Hamiltonians toward and away from integrable points, we show how integrability affects the behavior of EE's and PR's.  We explore in particular the statistical correlation between the EE's and PR's, quantifying the intuitive idea that one should be large in states where the other is large.  We elucidate how this relationship is affected by the location of the eigenstates in the spectrum, and by proximity to integrability.  We also show how both these quantities follow the ETH, in the quantitative sense that the eigenstate-to-eigenstate fluctuations of these quantities decrease with increasing system size as a power law with the Hilbert-space dimension $D$, similar to the eigenstate expectation values of local observables.

It is commonly thought that, in ground states and low-lying excited states of local Hamiltonians,
the entanglement entropy follows the so-called \emph{area law}~\cite{HolzheyEA1994, VidalEA_PRL2003,
  CalabreseCardy_JSTAT2004, Requardt_arXiv2006, Massanes_arealaw_PRA09,
  EisertEA2010,AlcarazSierra_PRL11}, with possible logarithmic corrections, while in the bulk of the
spectrum the entanglement entropy generally follows a \emph{volume law}.~\footnote{The two `laws' mean that the entropy of entanglement between two large spatial segments scales as the boundary between the segments, or as the size of the smaller segment, respectively.}  Reference~\cite{alba2009entanglement} has shown that, for an exactly solvable or integrable system, the bulk of the spectrum contains both area-law and volume-law eigenstates. Our results in this work shed further light on this picture.  In non-integrable cases, we show the bulk to have only `large' entanglement entropies, close to the values achieved by random wave functions. As one approaches integrability, eigenstates emerge in the bulk, that have low entanglement entropy, comparable to that of eigenstates at the bottom or the top of the spectrum. Loosely, one can identify the eigenstates with high and low entanglement with states having volume-law and area-law scalings, although unambiguous correspondence is not possible for the system sizes we are limited to with full numerical diagonalization.
Another interpretation of this characteristic behavior of both the PR and EE is that `typical' states in the bulk of the spectrum, i.e., those that correspond to thermodynamics, have large EE and PR, and that the appearance of low-EE and low-PR states in near-integrable cases reflects the existence of large numbers of non-typical states even in the middle of spectra for integrable systems.

Our comparative analysis shows that the overall distributions of EE's and PR's in many-body eigenstates are very similar. In non-integrable or generic systems, both quantities, when plotted against eigenenergies, are clustered near a sharply defined line of roughly `inverted-parabola' type, having small values at the edges of the spectrum and large values in the bulk of the spectrum. As one approaches integrability (either of the zero-interaction type or of the Bethe ansatz type), many states appear in the bulk which have lower values, so the scatter plot now takes a roughly filled-semicircle form. This appears to be a robust characterization of integrable versus non-integrable systems. An extension or application of this idea is seen in situations where the spectrum splits up into bands (e.g., at large interactions): the individual bands show one of the above two forms of scatter-plots, depending on whether or not the effective model describing the band is integrable. Looking at the statistical correlations between entanglement and PR's in all eigenstates, we find that there is positive correlation, but that this correlation is generally weaker in integrable systems because of the greater variance of both observables.  
We show that, in the non-integrable regime, the strength of fluctuations of both quantities from eigenstate to eigenstate scales as $D^{-1/2}$ with the Hilbert space dimension.  For the PR, this can be argued using the central limit theorem and the assumption of randomness of eigenstate
coefficients.  
Near integrability, the distributions of EE and PR values are not only wider than in the chaotic regime, but they are also skewed. We characterize the asymmetry of the distributions; our results point to a possible universality of such distributions in integrable models.  

Our results on the EE and PR as a function of eigenenergies provide a fundamental characterization of the distinction between integrable and non-integrable systems.  In particular, this is a way of understanding the lack of `usual' thermalization in integrable systems: even in the middle of the spectrum, integrable systems possess many eigenstates which are not `typical' and have EE's and PR's much lower than that expected of an effectively random state. The distribution of EE and PR values in the bulk of the spectrum characterizes the presence of non-typical states.  As is obvious from the `filled-semicircle' description above, the distributions have higher variance near integrability.  In addition, we find that, in the integrable region, the distributions have a definite skewness in the direction of smaller EE's and PR's.  The idea of non-typical states being characterized by low entanglement entropy is increasingly appreciated~\cite{alba2009entanglement, GroverFisher_JSTAT2014}.  Our results offer first steps toward a more complete future understanding of the emergence of non-typical states in integrable spectra, and brings to light the usefulness of participation measures in characterizing the appearance of such non-typical states.

We introduce in Sec.~\ref{sec_definitions} the definitions of the PR and EE and the tunable families of Hamiltonians we use.
We next discuss the dependence of EE's and PR's on eigenenergies in Sec.~\ref{sec_scatterplots}, identifying the forms of these scatter-plots that are characteristic for integrable, chaotic and decoupled regimes.  
Section~\ref{sec_correlations} quantifies the correlations between EE and PR using a correlation coefficient.  
Section~\ref{sec_fluctuations} provides a scaling analysis of the widths of the EE and PR distributions.  
Section~\ref{sec_skewness} provides analyses of the distributions of EE and PR values, in particular the skewness of these distributions, which at integrability characterize the presence of non-typical states with low entanglement.  
Section \ref{sec_discussion} provides some discussion and context.

\section{Definitions and models}
\label{sec_definitions}

\subsection{Participation ratio and Entanglement Entropy}

The participation ratio (PR) roughly measures how many basis states contribute to the eigenvectors. For any eigenstate $\ket{\psi_\alpha}$, we define it as
\begin{equation}
	\label{eqn_pr}
	P_\alpha = \left[D \sum_\gamma \abs{c^{(\alpha)}_\gamma}^4\right]^{-1},
\end{equation}
where $c^\alpha_\gamma = \braket{\phi_\gamma}{\psi_\alpha}$ are the eigenvector coefficients in terms of the basis $\{\ket{\phi_\gamma}\}$ and $D$ is the dimension of the Hilbert space. In the extreme case of only one $c^{(\alpha)}_{\gamma}$ being nonzero (the eigenstate being one of the basis states), the PR is $P_\alpha=1/D$.  In the other extreme case of all components being equal, i.e., $\abs{c^{(\alpha)}_\gamma}^2=1/D$, the PR is maximal, $P_\alpha = 1$.~\footnote{The convention/terminology varies in the literature.  Sometimes the factor $D$ is absent.  Often the inverse of the quantity~\eref{eqn_pr}, called the inverse participation ratio or IPR, is preferentially plotted.  Confusingly, sometimes the quantity~\eref{eqn_pr} with or without the factor $D$ is itself called the IPR~\cite{SantosRigol2010, RigolSantos2010, CanoviEA2011, BrandinoEA2012}, in contradiction to the literal meaning of the name.  The PR or IPR is sometimes also called the number of principal components or NPC~\cite{BrownEA2008, SantosRigol2010,  ZelevinskyEA_PhysRep1996, KotaEA2011}.}

The PR depends on the choice of the basis $\{\ket{\phi_\gamma}\}$. \emph{A priori}, there is no naturally preferred basis.  Since we are concerned with spatial correlations in this work, we will use basis states each of which is a fixed real-space configuration.  In any basis state, spatial regions of the lattice are not entangled with each other.
The PR in this basis is the natural many-body generalization of the PR widely used in single-particle physics (e.g., in the study of Anderson localization). 

The bipartite entanglement entropy (EE) is commonly used to study the strength of the quantum correlations between two parts of a many-body system. Given a partition of the system into parts A and B, the EE between A and B in the eigenstate  $\ket{\psi_\alpha}$ is 
\begin{equation}
	S_\alpha = -\Tr \rho_\mathrm{A}^{(\alpha)} \log \rho_\mathrm{A}^{(\alpha)} = -\sum_\gamma \lambda^{(\alpha)}_\gamma\log\lambda^{(\alpha)}_\gamma,
\end{equation}
where $\lambda^{(\alpha)}_\gamma$ are the eigenvalues of the reduced density matrix $\rho_\mathrm{A}^{(\alpha)} = \TrB \rho^{(\alpha)}$ of the $A$ part, obtained from the full density matrix $\rho^{(\alpha)}= \ket{\psi_\alpha}\bra{\psi_\alpha}$ of the eigenstate by tracing out $B$ degrees of freedom.

Unlike the PR, the EE is independent of the choice of basis.
The EE does however depend on the partition.  We consider partitions of real space, focusing on cases where $A$ and $B$ are connected blocks of sites.

\subsection{Tunable many-body Hamiltonians}

We use two model Hamiltonians, each with a tunable parameter $\lambda$ that allows to tune the vicinity to integrability. The Hamiltonians are of the form $H=H_0+\lambda H_1$, where $H_0$ is the Hamiltonian of an integrable model, and $H_1$ is a Hamiltonian in which the system is spatially decoupled and hence trivially integrable. Thus, in this setup each Hamiltonian family is non-integrable for intermediate $\lambda$ and integrable limits are approached for $\lambda\to0$ and $\lambda\to\infty$.

The first model is the \emph{Bose-Hubbard chain} of $L$ sites, with Hamiltonian
\begin{equation}
	H_{\mathrm{BH}} = -\sum_{i=1}^{L-1}(b_i^\dagger b_{i+1} +b_{i+1}^\dagger b_i) 
~+~ \sum_i  \lambda_i b_i^\dagger b_i^\dagger b_i b_i,
\end{equation}
where $b_i$, $b_i^\dagger$ are bosonic operators for site $i$, and $\lambda_i= \lambda(1+0.1\delta_{i1})$ is the interaction strength, which has an increased value at site $1$ in order to break reflection symmetry in the system; we wish to extract and present generic properties of many-body systems not dependent on any particular symmetries.  The interaction strength $\lambda$ plays the role of tuning parameter toward and away from the free-boson integrable point at $\lambda=0$ and the decoupled (product state, also integrable) limit at $\lambda\to\infty$.  We shift the interaction slightly at the first site in order to avoid reflection symmetry.  The number of bosons $\nb$ is conserved and so we can restrict to a single $\nb$ sector.  The dimension of the Hilbert space of the $(L,\nb)$ sector is equal to the binomial coefficient $D=\binom{L+\nb-1}{\nb}$.

The second model, introduced in Refs.~\cite{BeugelingEA2014PRE,BeugelingHaqueMoessner_offdiag_arXiv2014}, is an asymmetric Heisenberg XXZ ladder, with one leg possessing an extra site, as to break reflection symmetries in the system.
The Hamiltonian is given by $H_\mathrm{ladder}=H_\mathrm{legs} +\lambda H_\mathrm{rungs}$. Where $H_\mathrm{legs}$ and $H_\mathrm{rungs}$ are the sums over the (nearest-neighbor) leg and rung couplings, respectively, of the Heisenberg XXZ coupling, $h_{i,j}=\frac{1}{2}(S^+_iS^-_j + S^-_iS^+_j)+\Delta S^z_iS^z_j$. Here $S^z_i$ is the spin $z$ operator (we set $\hbar\equiv 1$), and $S^\pm=S^x_i\pm S^y_i$ are the raising and lowering operators. The anisotropy parameter $\Delta$ is fixed at $0.8$, away from the special values $0,\pm 1,\pm\infty$. For $\Delta\sim1$, the qualitative results are similar and only weakly dependent on the exact value of $\Delta$. 

The number $\nup$ of up spins is a conserved quantity. The analysis can therefore be constrained to a fixed-$\nup$ sector.  The $(L,\nup)$ sector has Hilbert space dimension $D=\binom{L}{\nup}$.

\section{Participation ratio and entanglement entropy as function of energy}
\label{sec_scatterplots}

\begin{figure}[t]
	\center\includegraphics[width=95mm]{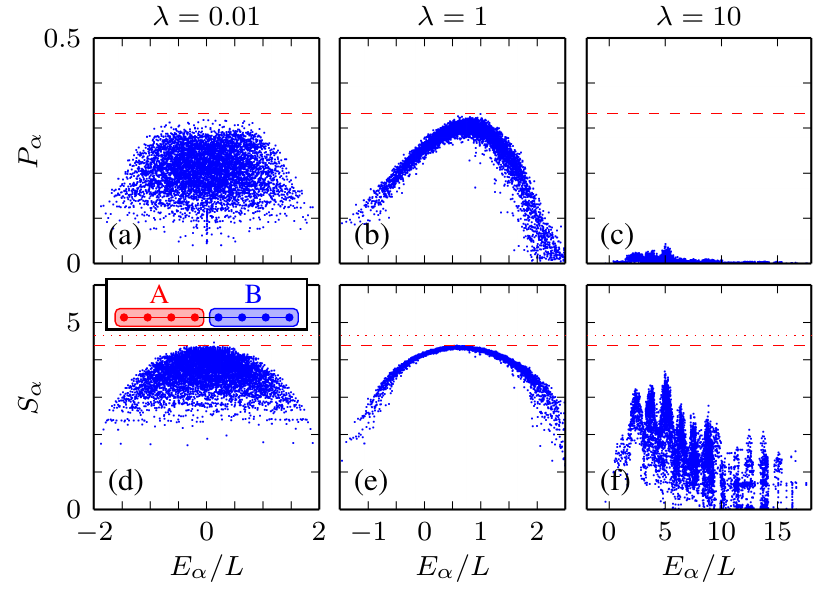}
	\caption{The participation ratio (PR) and entanglement entropy (EE) of the eigenstates for the Bose-Hubbard chain with system size $(L,\nb)=(8,8)$. The entanglement partition is between two equal halves of the chain [inset to (d)]. The integrability   (=interaction) parameter is $\lambda=0.01$ (a,d), $\lambda=1$ (b,e), and $\lambda=10$ (c,f). The values for random states are indicated by the red dashed lines, $P_\mathrm{rand}=1/3$ and $S_\mathrm{rand}=4.376 \pm 0.008$, respectively, and the maximal EE $S_\mathrm{max}=\log105$ is  indicated by the red dotted line.}
	\label{fig_pr_and_ee_bh}
\end{figure}

In \Fref{fig_pr_and_ee_bh} we plot the PR and EE of eigenstates as a function of eigenenergy for the Bose-Hubbard chain at unit filling ($\nb=L=8$), for $\lambda=0.01$, $1$, and $10$. These values are representative of the integrable (free-boson) limit $\lambda\to 0$, the intermediate non-integrable regime, and the integrable (decoupled) limit $\lambda\to\infty$. The EE is between partitions of equal halves of the chain. The Hilbert-space dimension of this system is $D=6435$.

In the small-$\lambda$ regime, the PR is low for eigenstates close to the spectral edges, but there are both low- and high-PR eigenstates in the bulk of the spectrum.
In the intermediate non-integrable regime, we observe a clear dependence on energy: The PR is low close to the edges of the spectrum, and the largest values $\sim 1/3$ are found in the bulk.  We note that the dependence of PR can be regarded as a smooth function of $E$ plus random fluctuations, i.e., the PR follows the ETH just like local observables.  This behavior will be further quantified in Section~\ref{sec_fluctuations}.

In the large-$\lambda$ (i.e., atomic) limit, the PR is much smaller. The reason is that the eigenstates are close to the basis states in this limit, because the hopping term in the Hamiltonian, which connects basis states, is negligible compared to the interaction ($\lambda$) term, which is diagonal in the Fock basis. As a result, the eigenstates are each dominated by a few basis states, i.e., the PR is small.
 
For the EE, we find roughly similar behavior as for the PR. In the non-integrable regime, the EE follows a smooth curve, with additional random fluctuations whose magnitudes decrease with increasing system size, consistent with the ETH~\cite{deutsch2013microscopic}. The shape of the curve --- large EE's in the spectral bulk and small EE's at the spectral edges --- suggests that the eigenstates have area-law character near the spectral edges and volume-law character in the spectral bulk. Near the free-boson limit (small $\lambda$), the EE is small near the spectral edges, while in the spectral bulk both small and large EE's are visible. Roughly speaking, the EE's of the states can be found in a range with a constant minimum and a maximum that depends smoothly on the energy. This suggests that, for integrable chains, the spectral edges contain eigenstates with area-law behavior, while the spectral bulk contains both area-law and volume-law eigenstates, consistent with the results of  Ref.~\cite{alba2009entanglement}.

In the large-$\lambda$ regime, the EE's are generally smaller, for the same reason as explained above for the PR. The decrease does not look as drastic as in case of the PR, presumably due to the logarithmic definition of the EE.

\begin{figure}[t]
	\center\includegraphics[width=95mm]{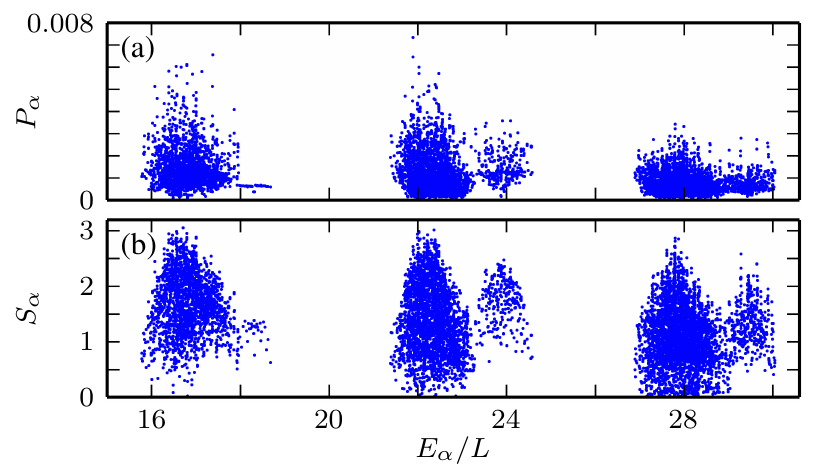}
	\caption{Magnification of the banded structure of (a) the PR and (b) the EE at large $\lambda$ for the Bose-Hubbard chain with $(L,\nb)=(9,9)$ and $\lambda=50$. Only a part of the spectrum is shown.}
	\label{fig_bh_large_lambda} 
\end{figure}

We compare the PR and EE values to values for `random' states (dashed horizontal lines), to check the idea that eigenstates in the spectral bulk are effectively or nearly random. We consider normalized vectors with random coefficients drawn from a Gaussian distribution with variance $1/D$, where $D$ is the Hilbert-space dimension. This distribution resembles that of the coefficients in the middle of the spectrum of typical non-integrable systems~\cite{Deutsch1991,NeuenhahnMarquardt2012}. For the PR, the random-state value $P_\mathrm{rand}$ is equal to $1/3$, which follows directly from the fourth moment of the Gaussian distribution. The random-state value $S_\mathrm{rand}$ of the EE is computed numerically using $10^4$ random states with Gaussian-distributed coefficients.

We also compare to the maximal possible values of the PR and EE. The maximal value of the PR is $1$ (corresponding to the case that the state is one of the basis states), far larger than the typical random-state value. For the EE, however, the maximal value, the logarithm of the Hilbert space dimension of the smaller partition, is only slightly larger than the random-state value. In the thermodynamic limit, $S_\mathrm{rand}/S_\mathrm{max}\to 1$~\cite{Page1993}.

\Fref{fig_pr_and_ee_bh} shows that, in the chaotic (non-integrable) regime, the EE and PR values in the middle of the spectrum are both very close to the random-state values. This is an important quantification of the widespread idea that the eigenstates at the middle of non-integrable spectra are effectively random for many purposes.
As the system size is increased, the EE and PR values are more sharply concentrated around a smooth line (c.f.\ Section~\ref{sec_fluctuations}), and this smooth line in increasingly close to the random state value. We conjecture that, in the thermodynamic limit, the EE and PR are increasingly concentrated near $P_\mathrm{rand}$ and $S_\mathrm{rand}$.  For finite sizes, the `smooth line' passes below the random values, but due to the fluctuations there are states with $P_\alpha>P_\mathrm{rand}$ and $S_\alpha>S_\mathrm{rand}$.  We have observed numerically that the fraction of states violating $P_\mathrm{rand}$ and $S_\mathrm{rand}$ as upper limits decreases with increasing system size.


In \Fref{fig_bh_large_lambda} we zoom into the large-$\lambda$ data. The EE for individual bands have forms which look more like the integrable case, i.e., a filled region rather than a relatively narrow curve. This suggests that the effective models that describe the dynamics within each energy sector at large $\lambda$ are themselves integrable models.  

\begin{figure}[t]
	\center\includegraphics[width=95mm]{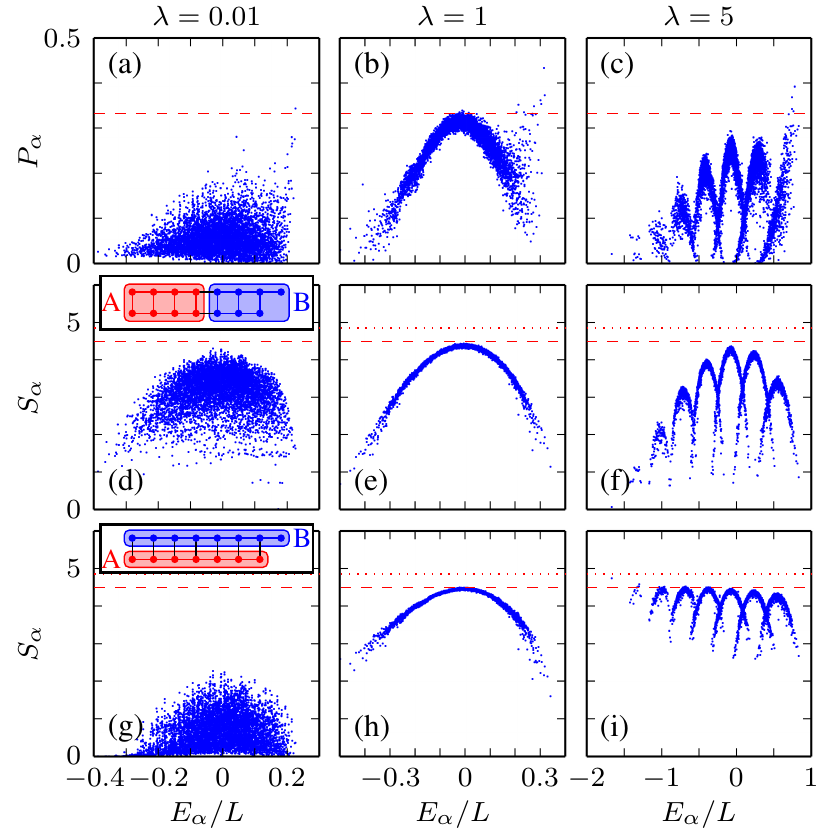}
	\caption{The eigenstate PR's, and the EE's for two different partitions, in the XXZ ladder $(L,\nup)=(15,7)$. Three different values of the integrability parameter (rung coupling) are used for each case.  (a)--(c) PR.  (d)--(f), EE's for a transverse partition. (g)--(i) EE's for a longitudinal partition, i.e, EE between the two legs.  Red dashed lines are random-state values: $P_\mathrm{rand}=1/3$ and $S_\mathrm{rand}= 4.488\pm0.010$. Red dotted line is maximal EE, $S_\mathrm{max}=8\log 2$.} 
	\label{fig_pr_and_ee_xxz}
\end{figure}

\Fref{fig_pr_and_ee_xxz} presents an analogous analysis for the XXZ ladder. The system size $(L,\nup)=(15,7)$ corresponds to the same Hilbert-space dimension $D=6435$ as the Bose-Hubbard example of \Fref{fig_pr_and_ee_bh}.  For the entanglement, we consider a transverse partition (cutting each leg into two) and a longitudinal partition (cutting each rung and measuring the entanglement between the two legs). The overall behavior is very similar to the Bose-Hubbard case. The two partitions give similar results for the EE, except for the small-$\lambda$ limit where the EE between the legs becomes small due to the legs being decoupled, as $\lambda$ is the rung coupling.

In the large-$\lambda$ limit, the spectrum has a banded structure. Loosely speaking, each band corresponds to a different number of triplet-like rungs and singlet-like rungs.~\footnote{The singlet/triplet language is inexact for $\Delta\neq1$, but for $\Delta=0.8$ it should still be a qualitatively useful description.} The fact that the PR and the EE's have an `inverted-parabola' shape within each band, indicates that the effective models describing each such energy sector (band) are non-integrable.  This is different from the bands of the Bose-Hubbard chain case shown in \Fref{fig_bh_large_lambda}, where each band individually has integrable forms.

As in the Bose-Hubbard case, near the center of the spectrum, the PR and EE reach the values close or even above the random-state expectation values.

\section{Correlations between participation ratio and entanglement entropy}
\label{sec_correlations}

\begin{figure}[t]
	\center\includegraphics[width=95mm]{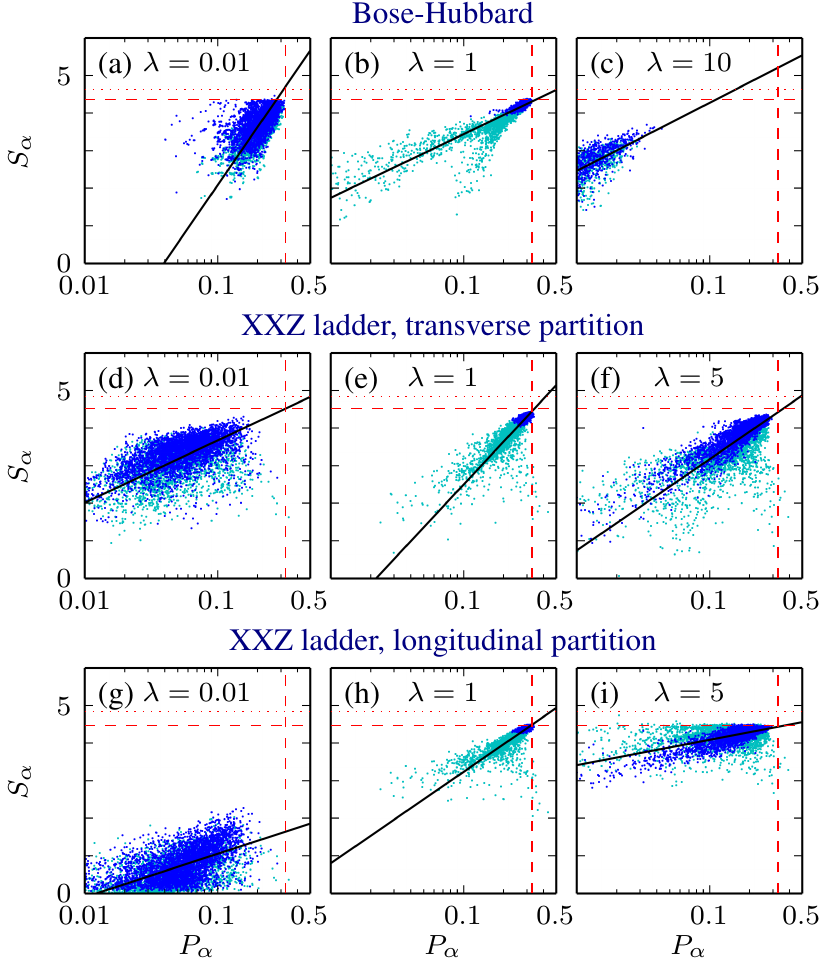}
	\caption{Scatter plots of PR versus EE.  The PR (horizontal axes) is plotted on a logarithmic scale. The dark blue points indicate eigenstates in the bulk of the spectrum, the light blue dots the states with the lowest $20\%$ and highest $20\%$ of  eigenvalues. The red dashed lines indicate the random-vector values $P_\mathrm{rand}$ and $S_\mathrm{rand}$, and the red dotted line $S_\mathrm{max}$. The black solid lines are linear fits obtained by the method of ODR.}
	\label{fig_correlations}
\end{figure}

The qualitatively similar behavior of the PR and the EE as function of energy, seen in the previous section, represents the idea that they probe physically similar properties of the eigenstates, namely, delocalization in configuration space. The intuitive expectation is that an eigenstate with high PR, i.e., composed of many basis states, has higher entanglement than a state with low PR. This motivates the study of statistical correlations between the two quantities, which we describe in this section.

In \Fref{fig_correlations}, we present scatter plots of the PR's and EE's of the eigenstates, for both the Bose-Hubbard and XXZ ladder system.  In view of the logarithm in the definition in the EE, it is natural to plot the PR on a logarithmic scale as well.  In order to distinguish the behavior of the spectral bulk from the behavior of the spectral edges, we use dark blue dots for the middle 60\% eigenstates  of the spectrum, and light blue dots for the lowest 20\% and highest 20\% eigenstates of the spectrum.

In the non-integrable regime, we see a clear difference between the natures of the spectral edge and spectral bulk. The bulk eigenstates have PR's and EE's both clustered near the random-state expectation values, while the eigenstates at the bottom and top of the spectrum have much smaller values for these quantities. In the near-integrable small-$\lambda$ regimes, the difference is less sharp as many bulk states also have PR and EE values much smaller than the random-state expectations. The scatter plots for large $\lambda$ show features similar to the small-$\lambda$ regime, although, as seen in the previous section, the situation is more complicated due to band structures, which are not obvious in the scatter plots.

Several system-specific features are also visible, e.g., the Bose-Hubbard chain case in the non-integrable regime, \Fref{fig_correlations}(b), shows a bifurcation of the data for the spectral edges. The two forks correspond to the top and the bottom of the spectrum.

The PR's and EE's clearly have positive statistical correlations in each case shown in \Fref{fig_correlations}.  The straight lines, obtained by orthogonal distance regression (ODR) fits to the full data sets, provide a visual idea of how well-correlated the two quantities are.

The quality of the statistical correlation can be quantified by a correlation coefficient. A linear dependence of random variables $\{(X_i,Y_i)\}$ is conveniently quantified by the Pearson product-moment correlation coefficient $r$, defined by 
\begin{equation}
	\label{eqn_pearson_corr}
	r^2 = \frac{\left[\sum_i(X_i-\bar{X})(Y_i-\bar{Y})\right]^2}{\sum_i(X_i-\bar{X})^2 \sum_i(Y_i-\bar{Y})^2}.
\end{equation}
In our case, $\log P$ and $S$ serve as $X$ and $Y$, respectively.

\begin{figure}[t]
	\center\includegraphics[width=95mm]{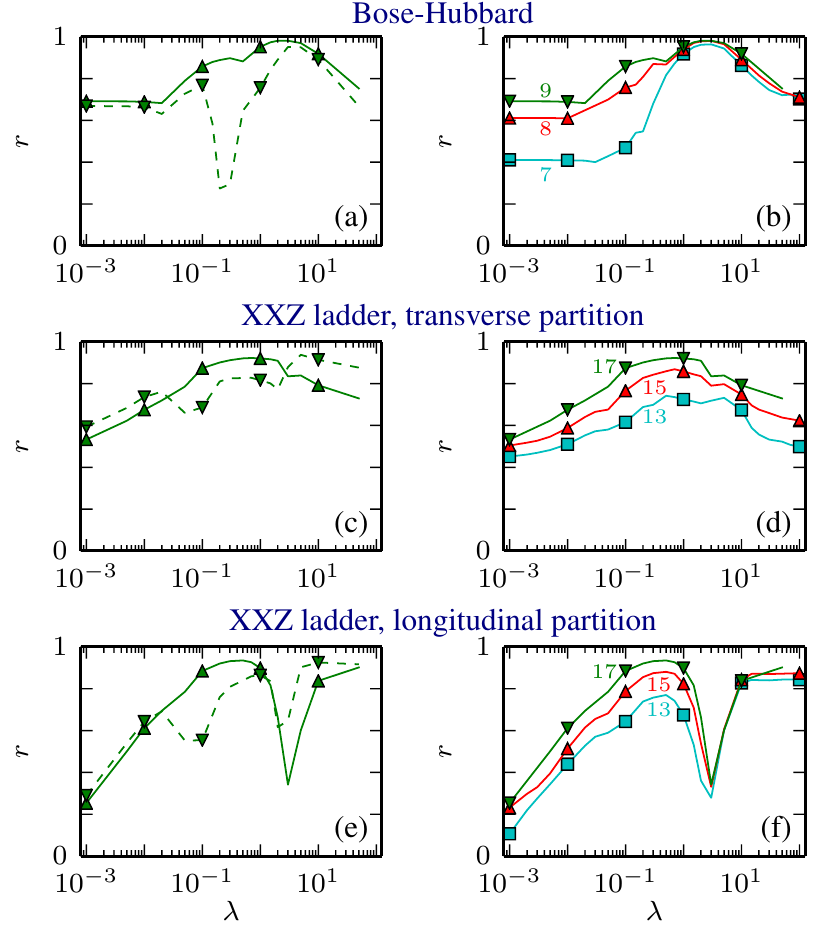}
	\caption{(a,c,e) The Pearson correlation coefficients $r$ of $\log P$ versus $S$ from all states (solid lines) and from bulk states (dashed lines) as a function of $\lambda$. The system sizes are $(L,\nb)=(9,9)$ for Bose-Hubbard and $(L,\nup)=(17,8)$ for XXZ   ladder. (b,d,f) Comparison of $r$ for different system sizes, with $L$ indicated near the curves. ($L=\nb$ for Bose-Hubbard chain; $L=2\nup+1$ for XXZ ladder.)}
	\label{fig_corr_coeff} 
\end{figure}

\Fref{fig_corr_coeff} shows Pearson coefficients $r$ of $\log P$ versus $S$ as a function of $\lambda$.  A recurrent feature is that the linearity coefficient has a maximum in the non-integrable regime and decreases as one approaches integrable points. This feature is common enough to be considered a generic characterization of integrable versus non-integrable systems; however, there are are several exceptions. In panels (a,c,e), the dashed lines correspond to the Pearson coefficients calculated from the bulk eigenstates only. These show a minimum in the highly non-integrable region.  The reason is that the bulk PR and EE are very clustered near the random-state expectation values; the spread of these values is not enough to show linearity.

Another exception is the case of entanglement between longitudinal partitions in the XXZ ladder. As $\lambda$ is increased, the coefficient $r$ increases again. This is because at large $\lambda$ the physics is more and more that of decoupled rungs. The entanglement between two sites of a single decoupled rung, is perfectly correlated with the PR of the two-site chain. The large-$\lambda$ behavior in this case should be regarded as two-site physics rather than the physics of integrability.

As there is no rigorous argument for the dependence between our variables ($S_\alpha$ and $\log{P}_\alpha$) to be linear, we have also quantified the correlation using the Spearman rank coefficient, which only measures the monotonicity of the functional dependence between $\{X_i\}$ and $\{Y_i\}$ without reference to linearity.  The overall behavior is very similar to the Pearson coefficient $r$, and hence not shown.

\section{Eigenstate-to-eigenstate fluctuations of the PR and the EE}
\label{sec_fluctuations}

\begin{figure}[t]
	\center\includegraphics[width=95mm]{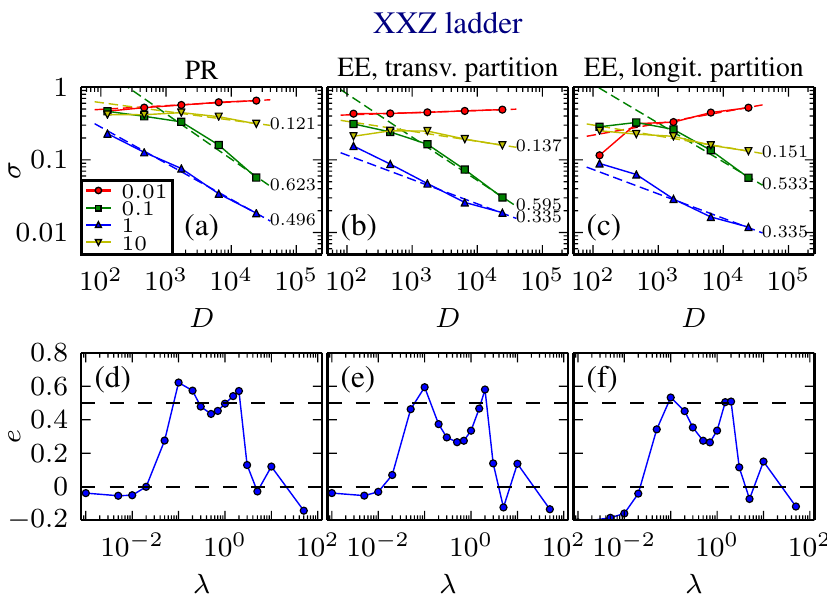}
	\caption{Fluctuation amplitudes of (a) the PR and (b,c) the EE as a function of Hilbert-space dimension $D$ for the XXZ ladder. The colors refer to different values of $\lambda$, see legend. We fit power laws $\propto D^{-e}$ to the data, and the exponent $e$ is shown on the right end of each fit. (d-f) The resulting exponents $e$ from the fits, as function of $\lambda$. In each case, five consecutive system sizes are used, with $L=2\nup+1$.}
	\label{fig_fluctuations}
\end{figure}

In this section, we quantify how the PR and EE follow the ETH, by studying the eigenstate-to-eigenstate fluctuations of these quantities as a function of system size, for the eigenstates in the middle of the spectrum. (The fluctuations are defined explicitly below.) This is the analog of the study of such fluctuations for local observables in Ref.~\cite{BeugelingEA2014PRE}.  Similar analysis has been performed using the Bethe ansatz for integrable models in Refs.~\cite{IkedaEA2013, Alba_arXiv1409}, also for local quantities.  The general result for \emph{local} quantities is that such fluctuations scale as $D^{-1/2}$ with the Hilbert-space dimension $D$ for non-integrable systems, i.e., exponentially with the system size, whereas they decrease as a power law with the system size for integrable systems. The generic non-integrable behavior $D^{-1/2}$ can be argued using the assumption that the eigenstate coefficients $c^{(\alpha)}_{\gamma}$ behave effectively as random variables for eigenstates in the middle of the spectrum~\cite{BeugelingEA2014PRE, Deutsch1991,NeuenhahnMarquardt2012}. 

For the fluctuations of the PR, we can construct a similar argument: the inverse quantity (IPR) can be written as $I_\alpha = P_\alpha^{-1} = \frac{1}{D} \sum_\gamma\left(D\abs{c^{(\alpha)}_\gamma}^2\right)^2$, an average of $D$ effectively random variables with unit variance.  The central limit theorem then predicts the strength of fluctuation in this quantity to scale as $\delta{I}\sim D^{-1/2}$.  Since $I_{\alpha}$ has nonzero average, its inverse $P_{\alpha}$ will have the same fluctuation scaling: $\abs{\delta{P}}\sim\abs{\delta{I}}/I^2 \sim D^{-1/2}$.  For the fluctuations of the EE, we do not have any simple argument for the scaling with $D$, since the entanglement is defined in terms of the eigenvalues of reduced density matrix, and we do not know much about the scaling properties or effective randomness of these numbers. References\ \cite{Deutsch2010, deutsch2013microscopic} provides arguments and data concerning the dependence of the EE fluctuations on the density of states.

Below, we provide numerical evidence that both the PR and EE fluctuations follow the $D^{-1/2}$ scaling in the non-integrable regime. 
We use a sequence of ladder system sizes with almost constant filling fraction: We choose $L=2p+1$ and $\nup=p$ for integer $p$, i.e.,  near-zero magnetization (near half filling).

As for local observables, the dependence of the PR and the EE as a function of eigenenergy $E_\alpha$ can be decomposed into a smooth component plus random fluctuations. In order to distinguish between the two, we define the smooth part as a moving average,
\begin{eqnarray}
	\label{eqn_moving_averages}
	\bar{P}(E) &\equiv \avg{P_\alpha}_{E_\alpha\in[E-\Delta E, E+\Delta E]}\\
	\bar{S}(E) &\equiv \avg{S_\alpha}_{E_\alpha\in[E-\Delta E, E+\Delta E]}
\end{eqnarray}
where the average is taken over all states with energies $E_\alpha$ within the interval $[E-\Delta{E}, E+\Delta{E}]$.  This is the `microcanonical' average~\cite{RigolEA2008,BeugelingEA2014PRE}. The interval width $\Delta E$ is taken as $0.05L$ as a compromise between good resolution and good statistics.

The measures of fluctuations $\sigma_{\Delta P}$ and $\sigma_{\Delta S}$ are then defined in terms of the variances 
\begin{equation}
	\label{eqn_fluctuations}
	\sigma_{\Delta P}^2 = \avg{(\Delta P_\alpha)^2} \quad\mbox{and}\quad
	\sigma_{\Delta S}^2 = \avg{(\Delta S_\alpha)^2},
\end{equation}
of the differences between the actual values and the moving averages, $\Delta P_\alpha\equiv P_\alpha - \bar{P}(E_\alpha)$ and $\Delta S_\alpha\equiv S_\alpha - \bar{S}(E_\alpha)$.  As in Ref.~\cite{BeugelingEA2014PRE}, the average $\avg{\cdots}$ above is taken with respect to the central $60\%$ of the eigenstates.

In \Fref{fig_fluctuations}(a--c) we show the dependence of the quantities $\sigma_{\Delta P}$ and $\sigma_{\Delta S}$ as a function of the Hilbert space dimension $D$, for the XXZ ladder.  As before, two types of partitions are used for the entanglement.  When a decreasing power law $\propto D^{-e}$ is found, the negative exponent $e$ is written to the right of the data set.
The data shows that in the intermediate regime, $\lambda=1$, the fluctuations of both PR and EE are strongly decreasing; the power-law analysis suggests that $\sigma_{\Delta P}\sim D^{-1/2}$ and $\sigma_{\Delta S}\sim D^{-1/2}$.  Close to the integrable limits, we observe a much slower decrease of the fluctuations, or even an increase.  

To evaluate the idea that the exponent is $-1/2$ in the non-integrable regime, we plot in
\Fref{fig_fluctuations}(d--f) the estimators for the exponent $e$ as a function of the tuning
parameter (rung coupling $\lambda$). There is a very clear overall trend that $e$ is close to $1/2$
for intermediate $\lambda$ and close to zero at small and large $\lambda$.  This behavior is very
similar to the scaling of the fluctuations of eigenstate expectation values of local observables
\cite{BeugelingEA2014PRE}.  A similar analysis of our Bose-Hubbard chain (not shown) shows the same
overall behavior.  Taken together, this is substantial numerical evidence in favor of $D^{-1/2}$
scaling of the fluctuations of these global quantities being generic for all non-integrable models.

At present, we do not understand the additional dip in the $e$ versus $\lambda$ curve appearing at
intermediate $\lambda$.  Also, a logarithmic correction cannot be completely ruled out from the
present data.~\footnote{It may be meaningful to check for logarithmic corrections for the EE
  fluctuations, because we do not currently have an independent argument for exact $D^{-1/2}$
  scaling of $\sigma_{\Delta S}$.}  However, the strong overall tendency for $e$ values to cluster
around $1/2$ for intermediate $\lambda$, and the similarity between the PR and EE cases, strongly
suggest $\sigma\sim D^{-1/2}$ behavior for both $P$ and $S$.

\section{Distribution of EE and PR:  skewness}
\label{sec_skewness}

\begin{figure}[t]
	\center\includegraphics[width=95mm]{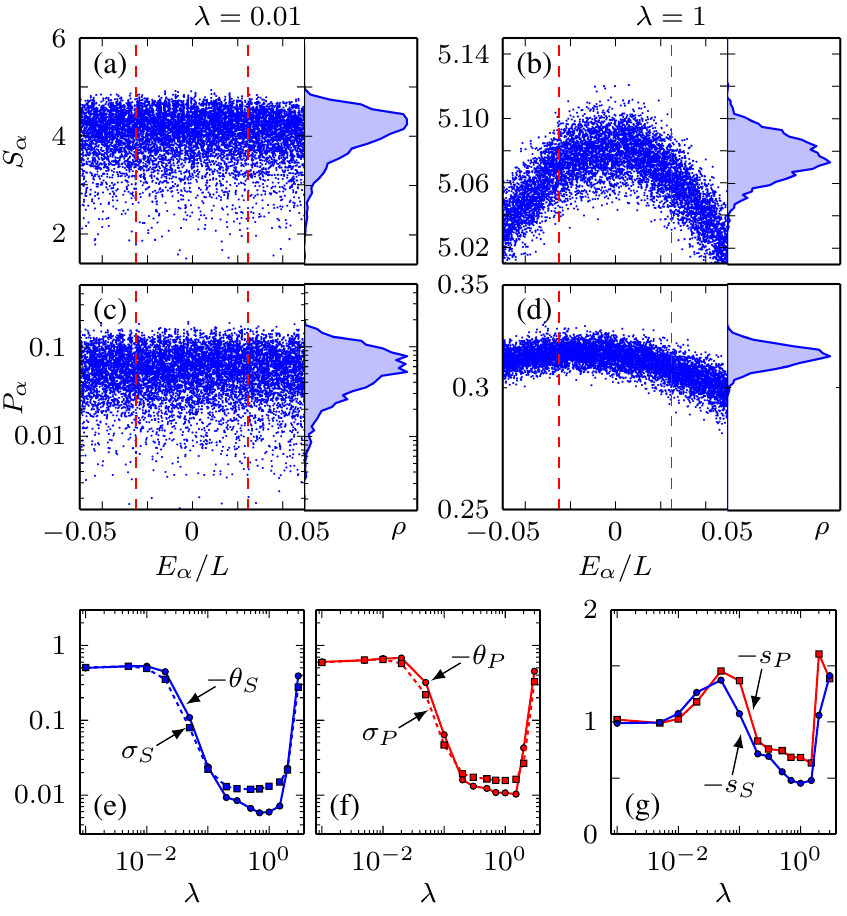}
	\caption{(a,b) EE values for transverse partition of the XXZ chain, close to the middle of the spectrum. For the eigenstates with energies between the vertical dashed lines, we plot the distribution of the $S_\alpha$. (c,d) The same for the logarithm of PR values, $\log(P_\alpha)$.  (e,f) Dependence on $\lambda$ of the standard deviation $\sigma$, and the negative cube root of the third moment $-\theta$, for the EE and the PR, respectively. (g) Cube root of the negative skewness $-s$ for the EE and PR as a function of $\lambda$.  System size: $(L,\nup)=(17,8)$.  The energy window used is $E_\alpha/L\in[-0.025,0.025]$.}
	\label{fig_skew}
\end{figure}

We have found the distributions of the EE and PR to give visibly different scatter plots in the integrable and the non-integrable regimes, e.g., in Figures~\ref{fig_pr_and_ee_bh} and~\ref{fig_pr_and_ee_xxz}. The distributions were characterized in terms of their widths in the previous section. In this section, we further characterize the distributions, focusing on their asymmetry or skewness.

In Figs.~\ref{fig_skew}(a)--(d), we show magnifications, zooming into the middle of the spectrum for the XXZ ladder. We plot the distributions for the EE and PR values in the small energy range $E_\alpha/L\in[-0.025,0.025]$. For the PR, the distribution of $\log P_\alpha$ (rather than $P_\alpha$ itself) is used.

In the chaotic regime, the distributions of the EE and the PR are near-Gaussian and near-symmetric, and have small width due to the $D^{-1/2}$ scaling in this regime. On the other hand, for small $\lambda$, the distributions are clearly skewed or asymmetric.
In order to quantify this asymmetry, we use the cube root of the third central moment of the distribution, $\theta_X = \avg{(X_\alpha-\avg{X_\alpha})^3}^{1/3}$, and compare this to the standard deviation $\sigma_X= \avg{(X_\alpha-\avg{X_\alpha})^2}^{1/2}$.  Here $X=S$ (EE) or $X=P$ (PR).  This comparison is displayed in Figs.~\ref{fig_skew}(e,f). 
In \Fref{fig_skew}(g), we quantify the shape of the distributions with the ratio $s_X=\theta_X/\sigma_X$, which is the cube root of the usual coefficient of skewness.

The EE and PR distributions are seen to have remarkably similar asymmetry properties. In the
non-integrable regime of $\lambda\sim 1$, both $\sigma_X$ and $-\theta_X$ are small, and their
relative magnitudes are such that the skewness is smaller than $1$.  The central limit theorem
(which is responsible for the $D^{-1/2}$ scaling of the distribution width of the PR and presumably
also the EE) predicts a Gaussian distribution, thus we expect $\theta_X$ and also $s_X=
\theta_X/\sigma_X$ to vanish in the large-size limit.

In the integrable regime, $\lambda\to 0$, the values of $-\theta_X$ and $\sigma_X$ are approximately
equal, and thus $s_x\approx{s_x^3}\approx1$.  We find the same behavior with the Bose-Hubbard chain,
where the $\lambda\to0$ limit corresponds to the integrable free-boson chain. This leads to the
conjecture that in integrable models the distributions of the EE and the PR may have a universal asymmetry with a skewness of unity.

In the $\lambda$ regime between the integrable ($\lambda\approx0$) and non-integrable ($\lambda\approx1$), the skewness rises above 1.  At present we do not know whether this effect disappears in the large-size limit.  Also, at larger $\lambda$ all the moments (and $s_X$) rise steeply for the energy window chosen; this effect is simply due to the band structure at large $\lambda$ and should be ignored.  
 
In the Bose-Hubbard model (not shown), qualitatively similar features are observed, including the near-unity skewness at small $\lambda$, and the non-monotonic behavior of $s_X$ as a function of $\lambda$.  The same behavior $-\theta_X \approx \sigma_X$ in two quite different integrable systems suggests that this could be a universal feature related to integrability.

\section{Discussion}
\label{sec_discussion}

We have provided a broad study of the behavior of the entanglement between roughly equal spatial partitions (measured using the EE), the delocalization in the space of real-space configurations (measured using the PR), and the statistical correlation between these two quantities, in all eigenstates of many-body Hamiltonians with local interactions.  We have used tunable Hamiltonians in order to identify characteristic features of integrable and non-integrable behaviors of the EE and the PR in the full eigenspectrum.

Our work adds to a growing literature on the study of full eigenspectra of local Hamiltonians, and in particular the study of entanglement and participation measures in all eigenstates. The block entanglement entropy in eigenstates far from the ground state has been studied for exactly solvable models in Refs.~\cite{alba2009entanglement, Alba_arXiv1409, MoelterAlba_arXiv1407, AresEA_JSTAT14, StormsSingh_PRE2014, LaiYang_arXiv1409}. Reference\ \cite{deutsch2013microscopic} has studied the entanglement entropy in all eigenstates in both generic and integrable systems (as we do here), in the context of understanding thermodynamic entropy in isolated quantum systems \cite{Deutsch2010, SantosEA2011, PopescuRohrlich_PRA97}.  Block entanglement entropies in eigenstates in the middle of the many-body spectrum are also of interest in the context of many-body localization~\cite{HuangMoore_arXiv1405, KjallBardarsonPollmann_PRL14}.
Data on PR or IPR in the full many-body spectrum of lattice systems have appeared previously in, e.g, Ref.~\cite{SantosRigol2010, RigolSantos2010}.
The information (Shannon) entropy, closely related to the PR, has been reported for the full eigenspectrum in Ref.~\cite{ZelevinskyEA_PhysRep1996} for the nuclear shell model, and in Refs.~\cite{SantosRigol2010, SantosEA2011} for lattice systems.

The correlation between the EE between spatial partitions and PR in the real-space configuration
basis is intuitively expected and has been mentioned in the literature, e.g., in
Refs.~\cite{GiraudEA2007, giraud2009entropy, SerbynPapicAbanin_PRL2013a, SerbynPapicAbanin_PRL2013b,
  LuitzEA_JSTAT2014, LuitzEA_PRB2014}.  Entanglement and participation measures have been compared
in ground states in Refs.~\cite{LuitzEA_JSTAT2014, LuitzEA_PRB2014}, and in random states in
Refs.~\cite{GiraudEA2007,giraud2009entropy}.  We have provided (Section \ref{sec_correlations}) an
explicit study of this correlation and a comparative study of the two quantities in all eigenstates
of clean systems, documenting explicitly the overall similar behavior of the two quantities.

By plotting the EE's and PR's as a function of eigenenergy, integrable and generic systems are found
to have characteristically different distributions of these quantities, a difference visually
explicit in our scatter plots (Section \ref{sec_scatterplots}).  The main difference is the
appearance of low-EE and low-PR states near integrability; this connects to the lack of usual
`typicality' in integrable systems, widely discussed in the context of thermalization.  The fact
that non-typical states are characterized by low entanglement is relatively well-known
\cite{alba2009entanglement, GroverFisher_JSTAT2014}; the present work shows that the same
distinction is also encapsulated by participation measures.  The EE and PR distributions are shown
in the chaotic regime to become smoother (narrower) at the same rate as the distributions of local
observables: the $D^{-1/2}$ scaling can be argued through randomness assumptions for the PR but is
more intricate for the EE. In the integrable regime, the distributions are skewed toward lower
values of EE or PR, due to the presence of the non-typical states.  For the two systems we have
studied, the skewness is near unity ($s_X\approx1$) in this regime.

The present work opens up several new questions for future investigations. 
First, the distributions at integrability should be studied more thoroughly in a larger class of
integrable and near-integrable systems.  While our results ($s_X\approx1$ for two systems) suggest
that there is some type of universality in the distributions, at present we have no explanation for
this behavior or understanding of the extent to which this is true for different integrable systems.
The system-size limitation of full diagonalization (and the requirement of good statistics within
small energy windows) also makes it difficult to judge the size-dependence of the distribution
asymmetry. Reference\ \cite{Alba_arXiv1409} has reported that the distribution of the two-site
entanglement entropy is Gaussian in large systems for the integrable XXX chain; this suggests that
the distribution of the EE of an $n$-site block might gradually acquire skewness as one increases
$n$ to about half the system size.  
Second, the $D^{-1/2}$ scaling of the EE presumably is rooted in the central limit theorem, but we
have not been able to formulate an explanation in these terms.  The difficulty lies in our lack of
knowledge about the scaling and distribution of the Schmidt coefficients (eigenvalues of the reduced
density matrix), when regarded as random variables.  Related analyses, connecting to the density of
states, appears in Ref.~\cite{Deutsch2010}.  
Third, one expects that the non-typical eigenstates (states with low EE and PR) are those where local observables deviate most from the microcanonical average value.  To the best of our knowledge, this has never been explicitly checked or quantified.  
Fourth, in the context of recent activity on many-body localization where the properties of the full eigenspectrum are commonly studied, a similar study of PR and EE distributions (perhaps as a function of proximity to the localization transition) might be of interest.
Finally, for free-fermion systems, Ref.~\cite{alba2009entanglement} provides an interpretation of the physical difference between non-typical and typical states.  A generic physical understanding for general near-integrable systems is currently lacking.  Conversely, one could also hope for a physical understanding of why the appearance of low-PR and low-EE eigenstates is suppressed in chaotic systems.

\ack
	The authors thank F.~Alet, J.~Bardarson, F.~Essler, R.~Moessner, F.~Pollmann, L.~Santos, and A.~Sharma for discussions.


\providecommand{\newblock}{}

\end{document}